\documentclass[12pt, authoryear, longmktitle]{elsarticle}

\usepackage[utf8]{inputenc}
\usepackage{
    amsfonts, amsmath, amssymb, array,
    booktabs,
    caption, color, comment,
    enumitem, eurosym,
    float, footmisc,
    geometry, graphicx,
    multirow,
    natbib,
    pdflscape,
    setspace, subcaption,
    threeparttable,
    ulem, 
}
\usepackage{hyperref}

\newtheorem{proposition}{Proposition}
\newtheorem{assumption}{Assumption}
\newtheorem{lemma}{Lemma}

\newtheorem{remark}{Remark}
\newenvironment{proof}[1][Proof]{\noindent\textbf{#1.} }{\hfill $\square$ \par}

\journal{}

\begin{document}


\begin{frontmatter}

    \title{Istanbul Flower Auction: The Need for Speed\tnoteref{t1}}
    \tnotetext[t1]{We are indebted to Sinan Ertemel for bringing this auction format to our attention and for his time and effort in his exploratory observations in Ayazaga, Istanbul. We also thank participants at various conferences, seminars, the Associate Editor, and three anonymous reviewers for their valuable suggestions.}

    \author{Isa Hafalir}
    \ead{isa.hafalir@uts.edu.au}

    \author{Onur Kesten}
    \ead{onur.kesten@sydney.edu.au}

    \author{Donglai Luo}
    \ead{donglai.luo@student.uts.edu.au}

    \author{Katerina Sherstyuk}
    \ead{katyas@hawaii.edu}

    \author{Cong Tao\corref{cor_auth}}
    \ead{cong.tao@student.uts.edu.au}

    \cortext[cor_auth]{Corresponding author. Business School, University of Technology Sydney, Sydney, Australia.}

    \begin{abstract}

        We examine a unique auction format used in the Istanbul flower market, which could transform into either Dutch or English auction depending on bidders' bidding behaviors. 
     By introducing a time cost that reduces the value of a perishable good as time passes,  
        we explore how this hybrid auction format accommodates the desire for speed via an adaptive starting price. We show that the Istanbul Flower Auction outperforms both the Dutch and English auctions in terms of the auctioneer's utility. With numerical analysis, we also illustrate the Istanbul Flower Auction's superiority in terms of social welfare and auction duration. Our results highlight the critical role of auction design in improving 
        welfare when the duration of the auction process plays a role.




    \end{abstract}


\end{frontmatter}


\section{Introduction} \label{sec:intro}

The study of auction mechanisms has long been a subject of interest in the field of economics, given their extensive application in various markets for the sale of goods and services \citep{milgrom1989auctions,klemperer1999auction}. Auctions are particularly vital in markets where goods need to be sold quickly and efficiently, such as those for perishable items that are auctioned off in numerous lots within a predetermined period. Fresh produce auctions, which deal with the sale of perishable goods like fruits, vegetables, fish, and flowers, are critical for ensuring these items reach consumers in a timely manner \citep{cassady1967}. These markets are characterized by their need for rapid sales to facilitate transactions between a large number of sellers and buyers within very tight time frames.

Among the various auction formats, the Dutch auction has been widely adopted in markets dealing with perishable goods, praised for its speed and efficiency. In this format, the auctioneer sets a high starting price, which is progressively lowered until a bid is made, ensuring a swift sale of the item at hand. Although it is not as common, the English auction is also used to sell perishable goods (for instance at the Tokyo fish market). In an English auction, the price starts low and gradually increases, with the object awarded to the highest bidder once no further offers are made. In comparison, the Istanbul Flower Auction introduces a nuanced approach to the auction mechanism by incorporating elements of both the Dutch and English auction formats, adapting either mechanism based on the initial responses of the bidders. Specifically, given a starting price $s$, the Istanbul Flower Auction operates as a Dutch auction if no one bids at $s$, and it operates as an English auction (among the initial bidders) if at least one bidder bids at $s$. This hybrid model is innovative in its flexibility, potentially offering a superior solution to the unique challenges faced in perishable goods markets.

In this paper, we present a comparative analysis of the Istanbul Flower Auction against the backdrop of traditional Dutch and English auctions. By focusing on the perishable goods market, particularly the sale of flowers in Istanbul, we explore how different auction formats can impact the outcomes and auctioneer's and bidders' utilities. Our theoretical framework centers on 
the shrinking value of a perishable good, modeling the time cost 
as a function of the total duration of the auction. Through this lens, we study the Istanbul Flower Auction, Dutch Auction, and English Auction in terms of the utilities of the auctioneer and bidders in perishable goods markets.

We establish that, whenever the time cost function is convex (which is a reasonably weak assumption), the Istanbul Flower Auction results in a strictly higher utility for the auctioneer compared to traditional Dutch and English auctions (Proposition \ref{prop::flower_better}). We also explore bidder welfare (Proposition \ref{prop::flower_bidder_better}), social welfare (Proposition \ref{prop::social_welfare}) and expected auction duration (Proposition \ref{prop::time-saving}) across different auction formats and establish that there exist starting prices that make Istanbul Flower Auction strictly superior to Dutch and English auctions. 

We further perform numerical analyses that illustrate  that for the starting price that maximizes the auctioneer's expected utility, the Istanbul Flower Auction performs better than Dutch auction in terms of social welfare, bidder welfare, and expected auction duration. 
The auction's performance improves significantly when bidders are impatient, offering higher utility for the auctioneer, greater bidder payoffs, and enhanced social welfare. While its relative benefits decrease with increased market competitiveness, it still performs favorably. The analysis also shows that alternative starting prices, optimized for bidder utility, social welfare, or auction speed, still benefit the auctioneer, underscoring the mechanism's robustness.

Our study is motivated by the premise that the choice of auction format can significantly affect market outcomes, especially in sectors where time is of the essence. By examining the Istanbul Flower Auction, we seek to contribute to the broader discourse on auction theory and practice, offering insights that could inform the design and selection of auction formats in various contexts. The findings of this research may have implications for auctioneers and market organizers worldwide, suggesting that flexibility in auction design could be key to optimizing sales processes, especially in markets where a large number of items need to be sold in very tight time frames.\footnote{These auctions are not only used for selling fresh produce. Used car auctions, which are used by dealerships or leasing companies, can sell hundreds to thousands of vehicles in a single event. Similarly, large wine auctions can involve the sale of thousands of bottles or cases, including rare and vintage wines, over the course of a single event.}

The rest of the paper is organized as follows.  Section~\ref{sec:litreview}  reviews  the related literature. Section \ref{sec:model} introduces our model and solves for the equilibrium of the Istanbul Flower Auction. Section \ref{sec:anal} provides theoretical results on the payoff comparison of the Istanbul Flower Auction against the backdrop of Dutch and English auctions. Section \ref{sec:num} provides our numerical results. Section \ref{sec:conclusion} concludes. We provide all proofs in the Appendix.

\section{Literature Review}
\label{sec:litreview}


There is a long strand of literature on the descending Dutch and the ascending English auctions, exploring various analogs and extensions derived from the standard models. Comprehensive overviews of these studies can be found in \cite{klemperer2004auctions} and \cite{milgrom2004putting}. Our contribution to the literature lies in studying a unique hybrid auction format that has been implemented in Turkish flower markets for decades.

Hybrid auctions typically allow for both ascending and descending prices throughout the auction process, with many variants in practice. For example, eBay provides a selling mechanism that gives bidders a ``buy-it-now" option simultaneously with the English auction process. \cite{mathews2004impact} shows that under such circumstances, impatient sellers would gain extra revenue by setting an attractive ``buy-it-now" price. Furthermore, \cite{azevedo2020channel} consider a descending buyout price and attribute its advantage over standard formats to information acquisition costs. In another example, \cite{katokroth2004} study multi-unit Dutch auctions for homogeneous goods with divisible lots where the price can drop and then rise, and explain it by synergy effects between parts of the lot.

Recent studies have recognized the importance of auction speed in the design of mechanisms for real-world markets. \cite{banks2003} document the trade-off between efficiency and speed in the Federal Communication Commission spectrum auctions, where speed and revenue can be enhanced through the improved design proposed by \cite{kwasnica2005}. \cite{andersson2013} numerically show that a hybrid Vickrey-English-Dutch algorithm is faster than the Vickrey-English or Vickrey-Dutch auctions. The role of speed in auctions has also been highlighted in experimental studies, where participants' impatience was linked to their enjoyment of participation \citep{cox1983} or intrinsic costs of time \citep{katok2008}.

However, none of these models formally considers ``time discounting'' in values due to auction duration, 
 except for \cite{hafalir2023speed}, which predates our work. 
\cite{hafalir2023speed} present 
a comprehensive analytical and experimental study of
a hybrid auction format used in various fish markets (including Honolulu and Sydney), and demonstrate 
how the participants' time costs affect the auction speed and the resulting participant welfare. Unlike the fish auctions in Honolulu and Sydney, in Istanbul Flower Auctions the ascending or descending price dynamics is determined by the initial decision of bidders and cannot be reversed later. 
More specifically, 
in Honolulu-Sydney fish auction 
the price can start going up after a bid in the descending Dutch phase, whereas this is not allowed in the Istanbul Flower Auction. In addition to the time-discounting modeling differences between our paper and \cite{hafalir2023speed}, one particular difference in the findings is that in \cite{hafalir2023speed}, the seller’s preference for a hybrid format depends sharply on the number of bidders (the hybrid auction is preferred by the auctioneer over the Dutch auction only when the number of bidders is small), whereas in the present paper, it does not.

\section{Model} \label{sec:model}

We consider a single-item auction with $n$ bidders in which the total time elapsed in the auction is important in the evaluation of the item for bidders. 

\subsection{Formal Description of the Istanbul Flower Auction}

The Istanbul Flower Auction proceeds as follows.\footnote{For a more detailed description, see \cite{oz2010alternative}.} It begins with a starting price announced by the auctioneer and bidders' decisions to {\it bid} or {\it wait} at the starting price. Then, the auction turns into one of the following two formats depending on the number of bidders bidding at the starting price:

\begin{enumerate}[label={(\arabic*)}]
    \item If no one bids at the starting price, the auction then operates as a Dutch auction: the price starts going down until someone bids (or the auction ends with no winner when the price drops to its minimum).
    \item If at least one bidder bids at the starting price, the auction then operates as an English auction for those initial bidders only: the price starts going up until the second last of the initial bidders leaves the auction (the rise of the price stops immediately at the starting price if there is only one bidder who bids at the starting price).
\end{enumerate}

\subsection{Incorporating Time Costs}

We assume that the private values of the bidders are independently and identically distributed according to a twice differentiable cumulative distribution function $F(\cdot)$ over $[0,1]$ with a corresponding density function $f(\cdot)$. As the clock ticks, the time cost is modeled as a decreasing and differentiable discount factor $c(t)$ on the item value, reflecting the perishability of the item. 

Specifically, if the auction ends when a bidder with value $v$ buys at a price $p$ after $t$ units of time have passed, the auctioneer's utility is equal to the revenue:
\begin{equation*}
    U_A(p) = p, 
\end{equation*}
and the winning bidder's utility is
\begin{equation*}
    U_B(p, v, t) = c(t) \cdot v - p,
\end{equation*}
where the time cost function $c(t)$ is a strictly decreasing function of time $t$.
\footnote{%
Such representation of the time cost underscores the perishable nature of the good, with more valuable items losing more value in a given time.
Note that in our model, the seller does not experience the time costs directly, although a shorter auction duration also benefits the seller through better preserving the value of the good.
In practice, perishable good auctions do not employ reserve prices, as delaying the sale until later is not an option. Indeed, the good would lose its freshness (if not its entire value) if is not sold ``here and now." 
}
More specifically, we have $c: [0,1] \rightarrow \mathbb{R}_{+}$,\footnote{Note that total duration of the auction is at most $1$.} where $c(0) = 1$; moreover, $c^{'}(t) < 0$ except when $c(t) = 1$ for all $t \geq 0$. We refer to the case $c(t) = 1$ as the case of ``no time cost.'' Note that $t$ denotes the duration of the auction, hence when the starting price is $s$ and the selling price is $p$, $t$ is given by $t:=|p-s|$. We assume that $c'(\cdot) > -1$, which ensures that the necessary first-order condition for the equilibrium (i.e., the differential Equation  \ref{eq::dutch_ode} below) has a valid solution. Bidders who lose the auction would get a utility of $0.$ 

\subsection{Equilibrium Behavior}

We begin our analysis with the bidder behavior. 
Bidders simultaneously decide on whether to bid at the starting price. We assume that these decisions are made instantly, without the implied time cost. Moreover, in equilibrium, the decision to bid also reflects each bidder's preference for the ascending or descending auction dynamics in a consistent way such that she will not join the other dynamics.

Consider a starting price $s \in [0, 1]$ and  bidders implementing symmetric increasing bidding strategies. They need to decide on (i) whether to bid or wait at the starting price and (ii) at what price to bid or leave when they are active bidders in the subsequent Dutch or English auction phases.

We first discuss the bidder decision to bid or wait at the starting price.
Intuitively, it is always a dominated strategy for a bidder with item value $v < s$ to bid at the opening, because she can never acquire a positive payoff in the subsequent English auction phase. Therefore, as we restrict our attention to the bidding strategy that is monotonically increasing in value, 
we focus on the case featuring a cutoff $p(s) \in [s, 1]$ such that the bidder only bids at the starting price if her private value is greater than or equal to $p(s)$, and waits otherwise. In the remainder of the analysis, we use the common ``first-order approach'' in auction theory literature, assuming the existence of a symmetric equilibrium with increasing bidding strategies. This equilibrium is characterized by the first-order conditions for the bidder optimization problems.

We now define several auxiliary functions that will be useful later. Let us denote the distribution of the highest of $n-1$ random variables identically and independently distributed according to $F(\cdot)$ by a cumulative distribution function $G(\cdot)$ with a corresponding density function $g(\cdot)$, i.e., $G(v) = F^{n - 1}(v)$. Let us also denote the joint density function of the highest value being $v$ and the second highest value being $x$ among $n$ random variables identically and independently distributed according to $F(\cdot)$ by $h(v, x)$, i.e., $h(v, x) = n(n - 1)f(v)f(x)F^{n - 2}(x)$.

In the subsequent Dutch auction phase, given the starting price $s$, let us denote the symmetric equilibrium bidding function by $b(v, s)$. Since this Dutch phase starts from price $s$ and not from $1,$ it could be possible that there is a cluster of bids at $s$: i.e., we may have $b(v,s)=s$ for all $v \in [\lambda(s),p(s)]$ for some $\lambda \in (0,p(s))$.
We assume ties break evenly for those bidders potentially choosing to cluster their bids at $s$. Lemma \ref{lemma::cutoff} shows that, in equilibrium, there will be no cluster bids at $s$, and the value cutoff $p(s)$ is determined by the starting price $s$ via the Dutch bidding function when $p(s) < 1$.

\begin{lemma} \label{lemma::cutoff}
    In equilibrium, the probability that a tie occurs in the Dutch auction phase is zero. Moreover, the unique solution to the cutoff value $p(s)$ as a function of the starting price $s$ is given by $b(p(s), s) = s$ when $p(s) < 1$.
\end{lemma}

A necessary condition for $b(v, s)$ to be a symmetric equilibrium strategy is that the bidder maximizes her payoff at her own value $v$ instead of any alternative $z$. For $v, z \le p(s)$, it is the Dutch auction problem (in the presence of time costs):
\begin{equation*}
    v = \mathop{\arg\max}_{z} G(z)[c(s - b(z, s))v - b(z, s)]
\end{equation*}
which leads to the following ordinary differential equation derived from the first-order condition
\begin{equation} \label{eq::dutch_ode}
    \frac{\partial b(v, s)}{\partial v} = \frac{g(v)}{G(v)}\frac{c(s - b(v, s))v - b(v, s)}{1 + c'(s - b(v, s))v}.
\end{equation}
Indeed, the Dutch phase bidding strategy $b(v,s)$ is defined by the differential equation \ref{eq::dutch_ode} \footnote{Since $c'(\cdot) > -1$, we have $1 + c'(s - b(v, s))v > 0.$} and the initial condition $b(0) = 0$. 

Now, consider a bidder with value $v$ who bids at the starting price to initiate the English auction phase, i.e., when $v \ge p(s)$. She will remain in the auction and compete with other bidders (if any) until the price reaches the level at which her utility drops to zero. Let us denote this English phase leaving strategy in the symmetric equilibrium by $m(v, s)$. It is defined by the following equation:
\begin{equation} \label{eq::english_strategy}
    c(m(v, s) - s)v - m(v, s) = 0.
\end{equation}
From the basic assumptions of the time cost function, $m(v, s)$ is a monotonic increasing function in the item value $v$.\footnote{Taking the derivative with respect to $v$ on both sides of the equation, we have $\frac{dm(v, s)}{dv} = \frac{c(m(v, s), s)}{1 - c^{'}{m(v, s) - s}} > 0$.} 
The expected utility for a bidder who bids  in the English phase is given by:\footnote{The superscript $FE$ corresponds to the flower auction in the English auction phase. Similarly, $FD$: flower auction in Dutch auction phase, $D$: Dutch auction, $E$: English auction, $F$: Flower auction.}
\begin{equation*}
    EU_{B}^{FE}(v, s) = (v - s)G(p(s)) + \int_{p(s)}^v [c(m(x, s) - s)v - m(x, s)] \, dG(x)  ,
\end{equation*}
where the first term is her expected utility from winning the item at the starting price when no other bidder is competing with her, and the second term is her expected utility from winning the item at the highest price among other competitors.

As such, when an Istanbul Flower Auction starts at $s$, the ex-ante expected utility for a bidder  is given by
\begin{multline} \label{eq::EUb}
    EU_{B}^{F}(s) = \int_0^{p(s)} [c(s - b(v, s))v - b(v, s)]G(v) \, dF(v) \\+ \int_{p(s)}^1 \left( (v - s)G(p(s)) + \int_{p(s)}^v [c(m(x, s) - s)v - m(x, s)] \, dG(x) \right) \, dF(v)
\end{multline}

The resulting utility for the auctioneer is given by
\begin{multline} \label{eq::EUa}
    EU_{A}^{F}(s) = \int_0^{p(s)} b(v, s) \, dF^n(v) \\+ \int_{p(s)}^1 \left( \int_0^{p(s)} sh(v, x) dx + \int_{p(s)}^v m(x, s)h(v, x) \, dx \right) \, dv
\end{multline}

Now, the optimization problem for the auctioneer is to select the starting price $s^*$  that maximizes her utility. That is,
\begin{equation*}
    s^* = \mathop{\arg\max}_{s} EU_{A}^{F}(s)
\end{equation*}

We can argue that the equilibrium we consider for the Istanbul Flower Auction is ``value-efficient.'' The reasons are that: (i) the Dutch phase bidding function is assumed to be increasing in the bidders' item values, and (ii) the English phase leaving function is increasing in the bidders' item values. 
The following remark notes the value-efficiency of the Istanbul Flower Auction.
\begin{remark}\label{rmk::efficiency}
    In equilibrium, Istanbul Flower Auction is value-efficient, in the sense that the item is allocated to the bidder with the highest item value.
\end{remark}

Note that due to the presence of time costs,  value-efficiency 
does not imply the social welfare maximization, as the latter includes both the auctioneer's and the bidders' utilities.
%
%
We investigate the social welfare and the duration of the Istanbul Flower Auction in addition to the auctioneer and bidder utilities. The corresponding definitions are given below.

The expected social welfare of the auction is defined as the aggregated expected utility for all market participants,  $EU_{S}^{F}(s) = EU_{A}^{F}(s) + n \cdot EU_{B}^{F}(s)$. We can calculate it as
\begin{multline} \label{eq::EUab}
    EU_{S}^{F}(s) = \int_0^{p(s)} c(s - b(v, s))v \, dF^n(v) \\+ \int_{p(s)}^1 \left( \int_0^{p(s)} vh(v, x) \, dx + \int_{p(s)}^v c(m(x, s) - s)vh(v, x) \, dx \right) \, dv.
\end{multline}

The expected duration of the auction is given by
\begin{multline} \label{eq::ED}
    ED^{F}(s) = \int _{0}^{p(s)} [s - b(v, s)] \, dF^n(v) \\+ \int _{p(s)}^{1} \left( \int_{0}^{p(s)} 0 \cdot h(v, x) \, dx + \int_{p(s)}^{v} [m(x, s) - s]h(v, x) \, dx \right) \, dv.
\end{multline}

\subsection{Dutch and English Auctions}

The Istanbul Flower Auction inherently contains elements of both English and Dutch auctions and can be explicitly converted to either format by setting the initial price at $s=0$ or $s=1$.

When $s=1$, as in the Dutch auction, the price continuously descends from $1$ until the first bidder stops the clock and wins the item. 
In this case, the ex-ante expected utilities for the auctioneer and the bidders, the social welfare and the expected duration are given by
\begin{align*}
    EU_{A}^{D} & = EU_{A}^{F}(1) = \int_{0}^{1} b(v, 1) \, dF^{n}(v),                  \\
    EU_{B}^{D} & = EU_{B}^{F}(1) = \int_{0}^{1} [c(1 - b(v, 1))v - b(v, 1)]G(v) \, dv, \\
    EU_{S}^{D} & = EU_{S}^{F}(1) = \int_{0}^{1} c(1 - b(v, 1))v \, dF^n(v),            \\
    ED^{D}     & = ED^{F}(1) = \int _{0}^{1} [1 - b(v, 1)] \, dF^n(v).
\end{align*}

The English auction is also incorporated into the Istanbul Flower Auction framework by setting the starting price at $s = 0$. In this case,
the aforementioned four auction characteristics are given by
\begin{align*}
    EU_{A}^{E} & = EU_{A}^{F}(0) = \int_{0}^{1} \int_{0}^{v} m(x, 0)h(v, x) \, dx \, dv,                \\
    EU_{B}^{E} & = EU_{B}^{F}(0) = \int_{0}^{1} \int_{0}^{v} [c(m(x, 0))v - m(0, s)] \, dG(x) \, dF(v), \\
    EU_{S}^{E} & = EU_{S}^{F}(0) = \int_{0}^{1} \int_{0}^{v} c(m(x, 0))vh(v, x) \, dx \, dv,            \\
    ED^{E}     & = ED^{F}(0) = \int_{0}^{1} \int_{0}^{v} m(x, 0)h(v, x) \, dx \, dv.
\end{align*}



Hence, the Istanbul Flower Auction can be viewed as a hybrid mechanism that is conducted as either an English or a Dutch auction. The initial price calibrates the occurrence of each format. 

\subsection{An Illustrative Example}

Consider the Istanbul Flower Auction with two bidders where the  time cost is $c(t) = 1 - 0.5t$. The private value of each bidder is independently and uniformly distributed in $[0,1]$. The auctioneer sets the starting price $s$.

The cutoff value $p(s)$ that sorts the bidders into those who bid and those who wait at the starting price is determined by equating the expected utility of the cutoff bidder in the Dutch phase and the English phase. If the auction starts as a Dutch auction, the Dutch bidding strategy $b(v, s)$ is determined by the differential equation below with the initial condition $b(0, s)=0$:
\begin{equation*}
    \dfrac{\partial b(v, s)}{\partial v} = \frac{(1 - 0.5s)v - (1 - 0.5v)b(v, s)}{v(1 - 0.5v)}.
\end{equation*}
The expected utility of the cutoff bidder with value $p$ in the Dutch phase is given by
\begin{equation*}
    EU_{B}^{FD}(p, s) = ([1 - 0.5(s - b(p, s)]p - b(p, s)) \cdot p.
\end{equation*}
Otherwise, the auction starts as an English auction, and the English leaving decision $m(v, s)$ can be solved from
\begin{equation*}
    [1 - 0.5(m(v, s) - s)]v - m(v, s) = 0 ,
\end{equation*}
which gives
\begin{equation*}
    m(v, s) = \frac{1 + 0.5s}{1 + 0.5v}v.
\end{equation*}
The expected utility of the cutoff bidder is given by
\begin{equation*}
    EU_{B}^{FE}(p, s) = (p - s) \cdot p.
\end{equation*}
We numerically solve $p$ from $EU_{B}^{FD}(p, s) = EU_{B}^{FE}(p, s)$ and observe that $p(s)$ increases from $0$ to $1$ in $s$ for $s \in (0, 0.557)$ and remains constant at $1$ for $s \in (0.557, 1)$.

By substituting in the functions $p(s)$, $b(v, s)$ and $m(v, s)$ calculated above, we can numerically find that the auctioneer's utility is maximized at the starting price $s^* = 0.462$. The cutoff value at the optimal starting price is $p(s^*) = 0.847$, which means that the bidders with item values $v \in [0, 0.847]$ will wait for the Dutch phase, while the bidders with item values $v \in (0.847, 1]$ will bid to start the English phase at the beginning of the auction. 
    
\begin{figure}
    \centering
    \includegraphics[width=0.7\linewidth]{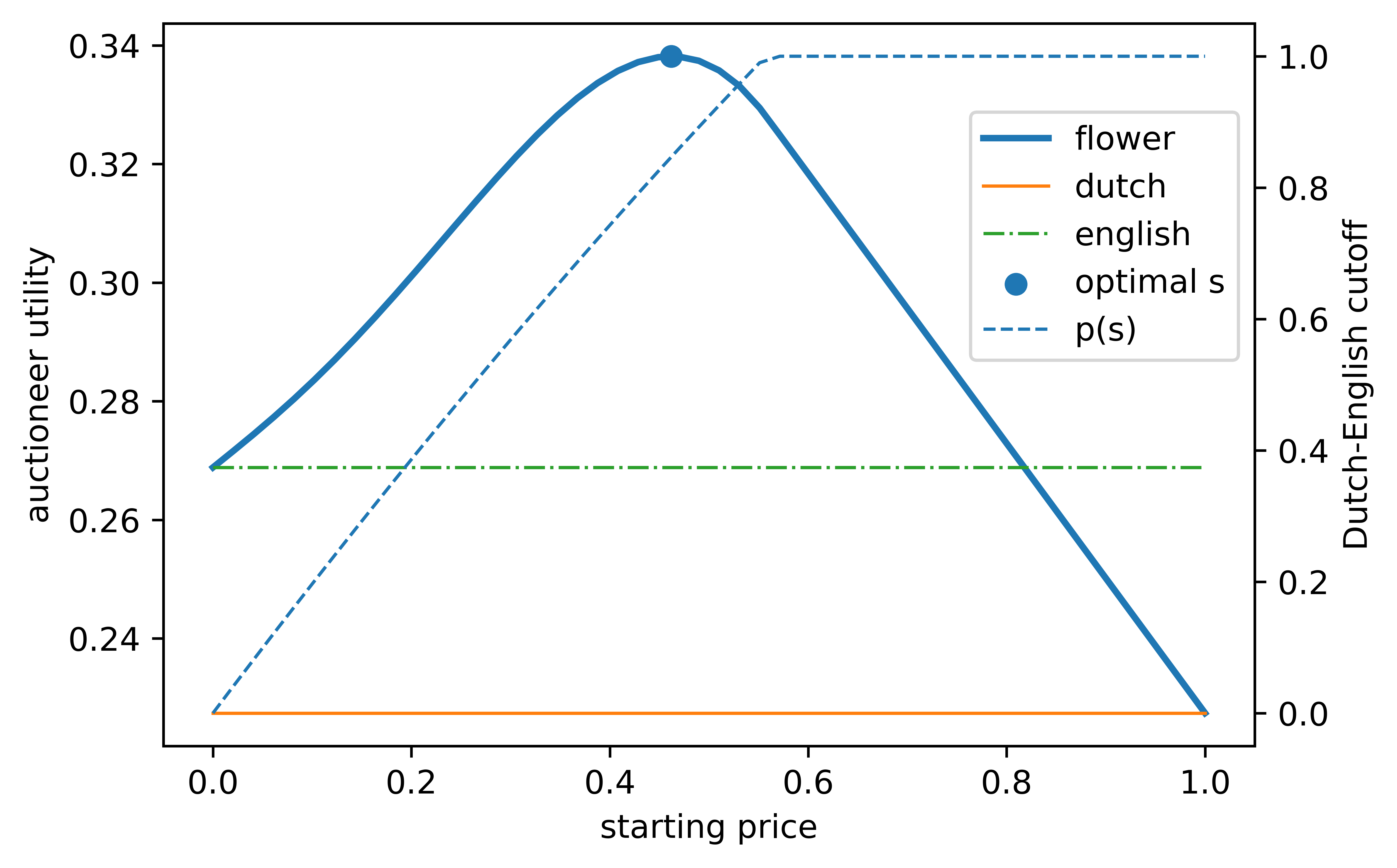}
    \caption{Expected utility of the auctioneer in two-bidder auctions with uniform value distribution and linear  cost of time $c(t) = 1 - 0.5t$}
    \label{fig:example_rev}
\end{figure}

Figure~\ref{fig:example_rev} compares the auctioneer's payoff in the Istanbul Flower Auction with the Dutch auction ($s=1$) and the English auction ($s=0$). The expected utility for the auctioneer at the optimal starting price is $0.338$, which is the peak of the utility curve. It outperforms the Dutch expected  utility of $0.227$ by $49\%$, and outperforms the English expected utility of $0.269$ by $26\%$. Yet notice that in this example the Istanbul Flower Auction can be worse than the English auction for the auctioneer if she sets a sufficiently high starting price. This illustrates the importance of setting the starting price optimally.

Under the optimal choice of the starting price, we can further calculate that the expected duration of the Istanbul Flower Auction is $0.138$, which is $82\%$ faster than the Dutch auction. 
Bidders are also expected to earn $58\%$ more than in the Dutch auction, which ultimately increases the social welfare by $53\%$.

\section{Auction Characteristics Analysis} \label{sec:anal}

\subsection{Superiority of the optimal Istanbul Flower Auction for the auctioneer}

When there is no time cost, the classic payoff equivalence result \citep{myerson1981optimal,riley1981optimal} can be obtained from Remark~\ref{rmk::efficiency}. We document it in the following proposition.

\begin{proposition} \label{prop:REP}
    The Istanbul Flower Auction satisfies the Payoff Equivalence Principle when bidders have no time costs, i.e., $EU_{A}^{F}(s) = \int_0^1 \left[ v - \frac{1 - F(v)}{f(v)} \right] dF^{n}(v)$ when $c(t) = 1$.
\end{proposition}

However, when the bidders incur time costs, the payoff equivalence principle no longer holds. In this more interesting case, we provide comparative static analyses based on the local properties of the auctioneer utility function in our time-costly setting.


The first-order derivative of the expected utility for the auctioneer in Istanbul Flower Auction  is given by
\begin{equation} \label{eq::dEUa}
    \begin{split}
        \frac{dEU_{A}^{F}(s)}{ds} = & - \frac{dp(s)}{ds}[s - b(p, s)]nF^{n-1}(p(s))f(p(s))                                     \\
                                    & - \frac{dp(s)}{ds}[m(p(s), s) - s]n(n-1)[1 - F(p(s))]F^{n-2}(p(s))f(p(s))                \\
                                    & + n[1 - F(p(s))]F^{n-1}(p(s))                                                            \\
                                    & + \int_{p(s)}^{1} \int_{p(s)}^{v} \frac{\partial m(x, s)}{\partial s}h(v, x) \, dx \, dv \\
                                    & + \int_{0}^{p(s)} \frac{\partial b(v, s)}{\partial s} \, dF^n(v).
    \end{split}
\end{equation}

Recall that $EU_{A}^{D} = EU_{A}^{F}(1)$. Hence, to demonstrate that the Istanbul Flower Auction is preferred by the auctioneer to the Dutch auction, it is sufficient to show that the expected utility of the auctioneer is decreasing at $s=1$ (or the derivative of the expected utility at $s=1$ is negative), since then it must reach a higher utility for some interior $s$. Similarly, since $EU_{A}^{E} = EU_{A}^{F}(0)$,  the Istanbul Flower Auction is better than the English auction if the expected utility for the auctioneer is increasing at $s=0$ (or the derivative of the expected utility at $s=0$ is positive). That is, the payoff comparison between the auction formats follows naturally from  establishing the sign of the derivative of the auctioneer's expected utility  with respect to the starting price at the corresponding boundary values.

It is evident from equation ~\ref{eq::dEUa} 
above
that the properties of the terms $\frac{\partial b(v, s)}{\partial s}$, $\frac{\partial m(v, s)}{\partial s}$ and $\frac{dp(s)}{ds}$ are crucial in determining the sign of the derivative.

We first note that when varying the starting price $s$, the variation in the Dutch phase bidder strategy $\frac{\partial b(v, s)}{\partial s}$ has an upper bound, and the variation in the English phase bidder strategy $\frac{\partial m(v, s)}{\partial s}$ is bounded:

\begin{lemma} \label{lemma::partial_b_s_upperbound}
    $\frac{\partial b(v, s)}{\partial s} < 1$.
\end{lemma}

\begin{lemma} \label{lemma::partial_m_s_bound}
    $0 \le \frac{\partial m(v, s)}{\partial s} < 1$, and the equality only holds when there is no time cost, i.e. $c(\cdot) \equiv 1$.
\end{lemma}

Intuitively, the variation of the starting price is only partially reflected in the variation of the Dutch-phase or English-phase bidding strategy.

Next, we describe how the bidder Dutch-English cutoff value $p(s)$ in the Istanbul Flower Auction changes with the starting price $s$ set by the auctioneer:

\begin{lemma} \label{lemma::increasing_cutoff}
    There exists a starting price $\tilde{s} \in (0, 1)$ such that
    \begin{enumerate}[label={(\alph*)}]
        \item $p(s) < 1$ and $\frac{dp(s)}{ds} > 0$ when $s \in [0, \tilde{s})$;
        \item $p(s) = 1$ and $\frac{dp(s)}{ds} = 0$ when $s \in [\tilde{s}, 1]$.
    \end{enumerate}
\end{lemma}

The bidder cutoff value $p(s)$ is monotonically increasing in the starting price as long as the starting price is below a certain value $\tilde{s}$; in this case, the
Istanbul Flower Auction can indeed proceed into either the Dutch or the English phase. 
If the starting price is set at or above $\tilde{s}$,
the Dutch-English bidder cutoff value is fixed at $1$.  
In that case, 
all bidders will wait for the Dutch phase, and the Istanbul Flower Auction simply becomes a Dutch auction with a starting price $s$.

This lemma also helps us conceptualize the 
``non-triviality" of the Istanbul Flower Auction dynamics from the perspective of the auctioneer who is 
choosing the
starting price among all possible $s \in [0, 1]$: (i) the auction is {\it non-trivial}  if $s \in (0, \tilde{s})$, in which case different bidders may differ in their decisions to wait for the Dutch phase or to bid in the English phase, 
depending on their value;  while (ii) the auction is {\it trivial}  if either $s=0$, in which case all bidders bid in the English phase, or if $s \in [\tilde{s}, 1]$, in which case all bidders wait for the Dutch phase.

So far, we have only established the upper bound of $1$ for $\frac{\partial b(v, s)}{\partial s}$, which is insufficient for further analysis. Further insights into how the Dutch phase bidding function reacts to changes in the starting price can be obtained by imposing a mild assumption on the time cost function:

\begin{assumption} \label{assumption::cost}
    The time cost function $c(t)$ is convex in its domain, i.e., $c^{''}(t) \ge 0$ for all $t \in [0,1]$.
\end{assumption}

Note that this assumption is satisfied for many standard time-cost functions used in the literature. Specifically, the exponential cost $c_{B}(t) = e^{- \mu t}$, the hyperbolic cost $c_{B}(t) = \frac{1}{1 + \mu t}$, and the linear cost $c_{B}(t) = 1 - \mu t$ all satisfy this assumption when the parameter $\mu \in [0, 1)$. 

The convexity of the time cost function ensures a negative sign of $\frac{\partial b(v, s)}{\partial s}$, which is given by the lemma below:

\begin{lemma}\label{lemma::dbds}
    In equilibrium, the bidder's Dutch phase bidding function $b(v, s)$ in the Istanbul Flower Auction is strictly decreasing in the starting price $s$ when there is a time cost and Assumption~\ref{assumption::cost} is satisfied. That is, $\forall v \in (0, p(s)), \frac{\partial b(v, s)}{\partial s} < 0$ when $c^{'}(t) < 0, c^{''}(t) \ge 0$. In addition, $\frac{\partial b(0, s)}{\partial s} \equiv 0$.
\end{lemma}

With the assumption and the lemmas introduced above, the payoff comparison between the Istanbul Flower Auction and the Dutch and the English auctions can be established. We now state our primary analytical finding on the general advantage of the Istanbul Flower Auction.

\begin{proposition} \label{prop::flower_better}
    When there is a time cost and Assumption~\ref{assumption::cost} is satisfied, the optimal Istanbul Flower Auction is strictly better than both the Dutch auction and the English auction for the auctioneer, and the resulting auction dynamics is non-trivial. That is, 
    if $c^{'}(t) < 0$ and $c^{''}(t) \ge 0$, then     $s^{*} \in (0, \tilde{s})$, and $EU_{A}^{F}(s^*) > EU_{A}^{D}$,   $EU_{A}^{F}(s^*) > EU_{A}^{E}$. 
\end{proposition}

\subsection{Possible improvements by the Istanbul Flower Auction from other perspectives}
\label{sec:sw}

Recall that the aforementioned ``optimal" Istanbul Flower Auction is derived from the perspective of the auctioneer, and hence the most preferred
starting prices for other parties may be different from the one chosen by the auctioneer. The three propositions below confirm the existence of a 
non-trivial 
Istanbul Flower Auction that strictly outperform both the Dutch auction and the English auction when the focus of optimization shifts to the bidder utility, the social welfare or the speed of the auction, respectively. To emphasize the difference from the auctioneer's optimal starting price $s^{*}$, the Istanbul Flower Auction starting prices for these improvements are denoted by $\hat{s}$. 

\begin{proposition} \label{prop::flower_bidder_better}
    When there is a time cost and Assumption~\ref{assumption::cost} is satisfied, there exists a starting price $\hat{s}_{B}$ such that the corresponding Istanbul Flower Auction is strictly better than both the Dutch and the  English auction for the bidders, and the resulting auction dynamics is non-trivial. That is,   $\exists \hat{s}_{B} \in (0, \tilde{s})$  such that $EU_{B}^{F}(\hat{s}_{B}) > \max \{ EU_{B}^{D}, EU_{B}^{E} \}$.
\end{proposition}

\begin{proposition}
    \label{prop::social_welfare}
    When there is a time cost and Assumption~\ref{assumption::cost} is satisfied, there exists a starting price $\hat{s}_{S}$ such that the  corresponding  Istanbul Flower Auction is strictly better than both the Dutch and the  English auction in terms of the social welfare,  and the resulting auction dynamics is non-trivial. That is, $\exists \hat{s}_{S} \in (0, \tilde{s})$  such that $EU_{S}^{F}(\hat{s}_{S}) > \max \{ EU_{S}^{D}, EU_{S}^{E} \}$.
\end{proposition}

\begin{proposition}\label{prop::time-saving}
    When there is a time cost and Assumption~\ref{assumption::cost} is satisfied, there exists a starting price $\hat{s}_{D}$ such that the corresponding  Istanbul Flower Auction is strictly better than both the Dutch and the English auction in terms of the auction duration, and the resulting auction dynamics is non-trivial. That is,  $\exists \hat{s}_{D} \in (0, \tilde{s})$  such that $ED^{F}(\hat{s}_{D})< \min \{ ED^{D}, ED^{E} \}$.
\end{proposition}


The proofs of the above propositions follow the same logic that we used to prove the superiority of the optimal Istanbul Flower Auction for the auctioneer by demonstrating that (1) the target function is increasing at the trivial English-equivalent starting price $s=0$, and (2) the target function is decreasing in the trivial Dutch-equivalent starting price interval $s \in (\tilde{s}, 1)$. 

The proof methods of the above propositions also establish a ``win-win'' scenario in the following sense. The Istanbul Flower Auction with a starting price slightly less than $1$ would result in a higher auctioneer utility and bidder (ex-ante) utility, as well as a lower expected duration, as compared to the Dutch auction. Similarly, the Istanbul Flower Auction with a starting price slightly greater than zero would result in a higher auctioneer utility and bidder (ex-ante) utility, as well as a lower expected duration, as compared to the English auction.

Deriving analytical results at the auctioneer-optimal starting price $s^{*}$ for these auction characteristics proves challenging. 
As a substitute, we present supportive computational results in the next section. Our numerical findings illustrate that, under a wide range of parameter choices, the auctioneer-optimal Istanbul Flower Auction is also a better choice than the Dutch auction for the bidders and the social planner, and it is faster.

\section{Numerical Analysis} \label{sec:num}

In this section, we provide further evidence for the advantages of the Istanbul Flower Auction by conducting numerical simulations for the model. Although our analysis of auction characteristics confirms the advantages of the Istanbul Flower Auction over both standard auction formats, we focus on the comparison with the Dutch auction in our numerical analysis for a number of reasons. Firstly, the Dutch auction is the most commonly used auction format for perishable goods in 
markets around the world, while the use of the English auction is less frequently observed.\footnote{There is a long history of the dominant use of the Dutch auction in the sale of fresh flowers. In fact, the name of the Dutch auction comes from the famous tulip auction in 17th-century Holland. However, for other perishable goods,  English (ascending) auctions are also observed in some contexts, for example, in the Tokyo fish market.} Secondly, the analysis process for the English auction will closely follow the approach used for comparing it with the Dutch auction. The main information conveyed in this section is that the Istanbul Flower Auction can bring significant benefits when implemented in real-world perishable goods markets where the Dutch auction is in place and many practical concerns are taken into account. Specifically, 
in this section
we explore the answers to the following questions: 

\begin{itemize}
    \item How will the auctioneer's benefit in the optimal Istanbul Flower Auction vary with the changes in the time cost parameter, and with market competitiveness (i.e., the number of bidders)?
    \item Will the auctioneer-optimal Istanbul Flower Auction enhance the benefits for others in terms of bidder utility, social welfare, and auction speed?
    \item If the starting price is set to enhance other aspects of the Istanbul Flower Auction, rather than prioritize auctioneer benefits, will the auction still serve the auctioneer's interests?
\end{itemize}

As we have noted before, typical functional forms for time costs, such as linear, exponential, and hyperbolic costs, all satisfy Assumption~\ref{assumption::cost}. For simplicity, we consider the linear time cost function $c(t) = 1 - \mu t$, with the parameter $\mu \ge 0$ indicating the impatience of bidders, and assume a uniform distribution of  item values $F(v) = v$ over $[0, 1]$.

\begin{figure}[!tbh]
    \begin{subfigure}{0.48\textwidth}
        \centering
        \includegraphics[width=\linewidth]{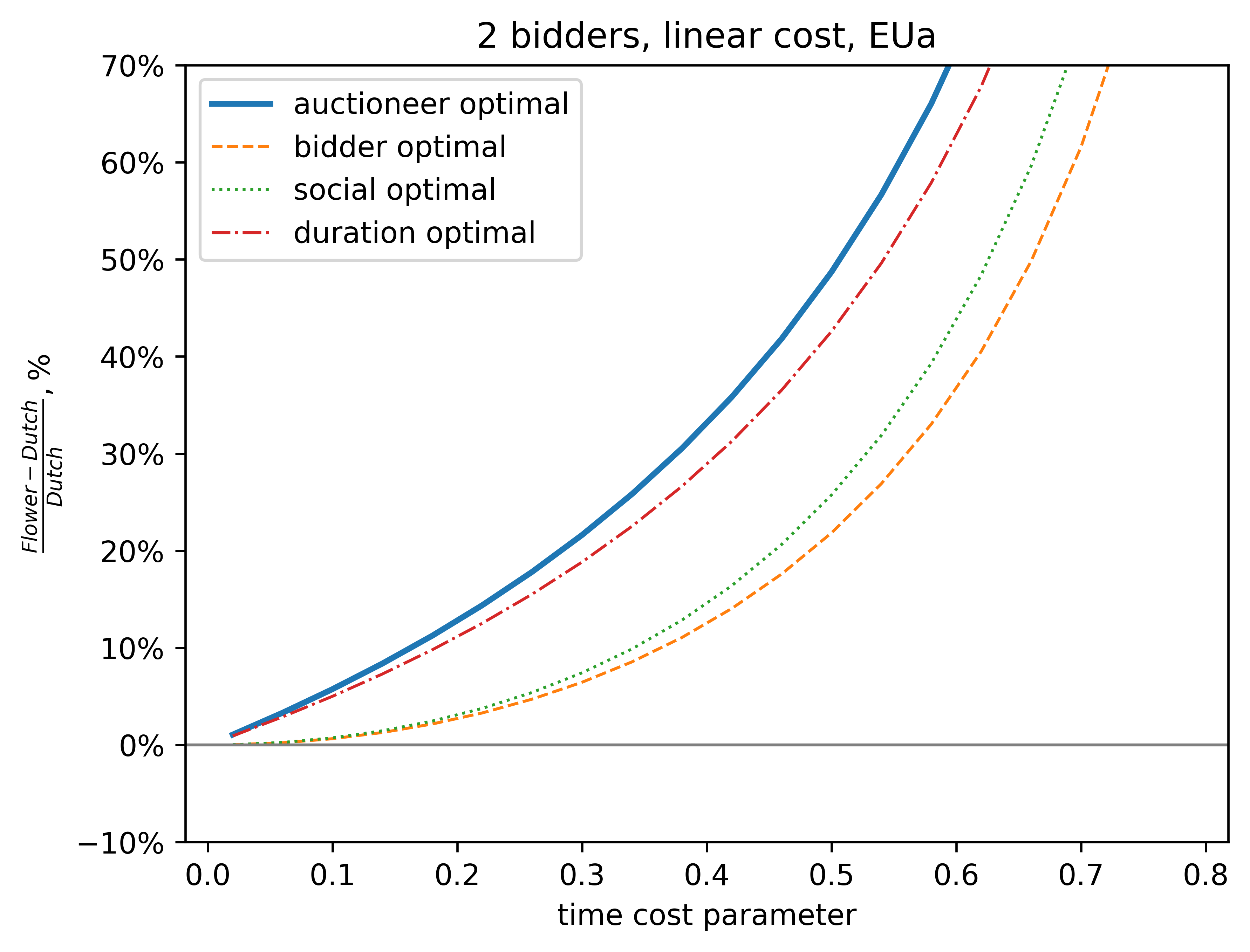}
        \caption{2 bidders}
    \end{subfigure}
    \begin{subfigure}{0.48\textwidth}
        \centering
        \includegraphics[width=\linewidth]{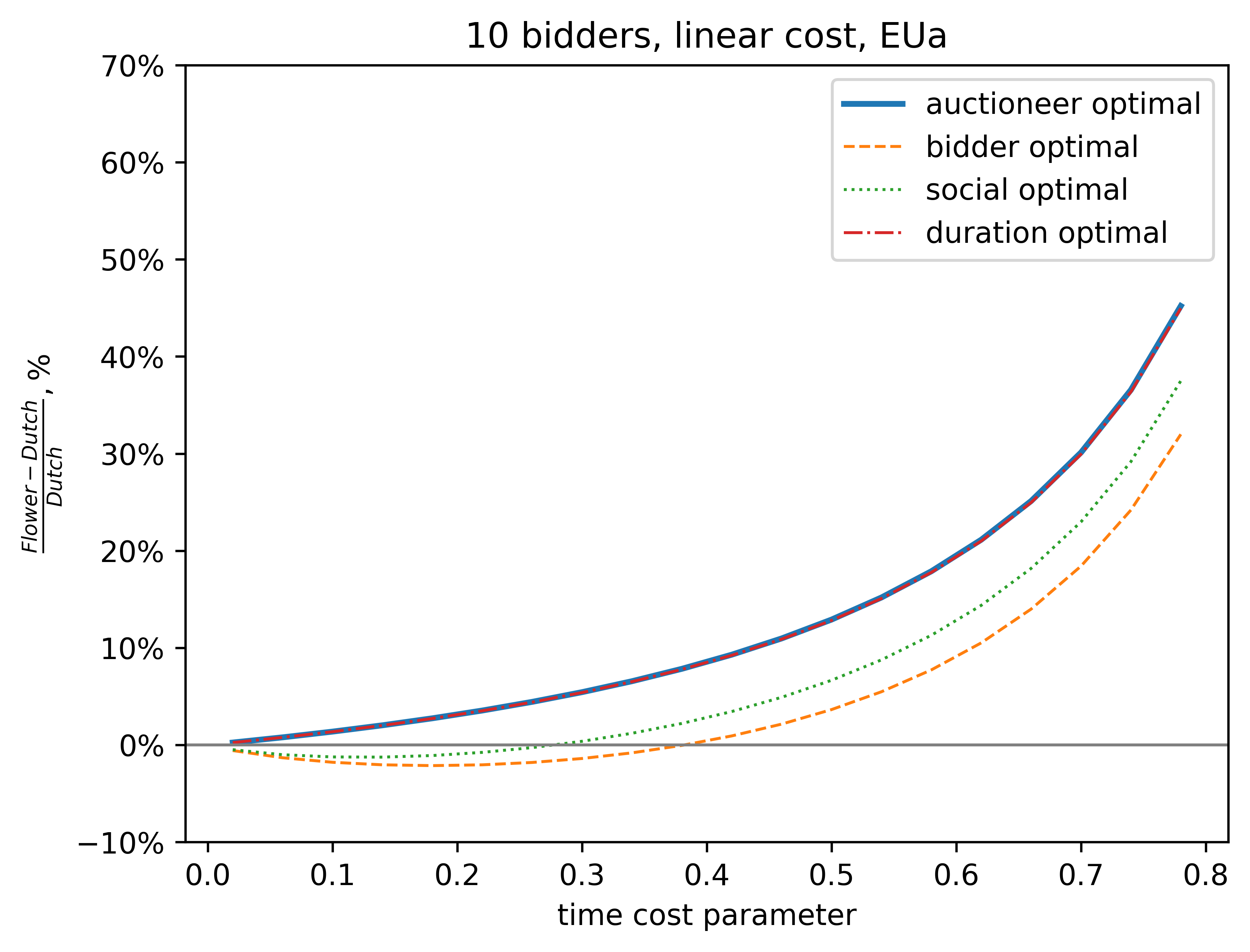}
        \caption{10 bidders}
    \end{subfigure}
    \centering
    \caption{Auctioneer utility under varying time cost}
    \label{fig:optimals_vary_cost_EUa}
\end{figure}

\begin{figure}[!tbh]
    \begin{subfigure}{0.48\textwidth}
        \centering
        \includegraphics[width=\linewidth]{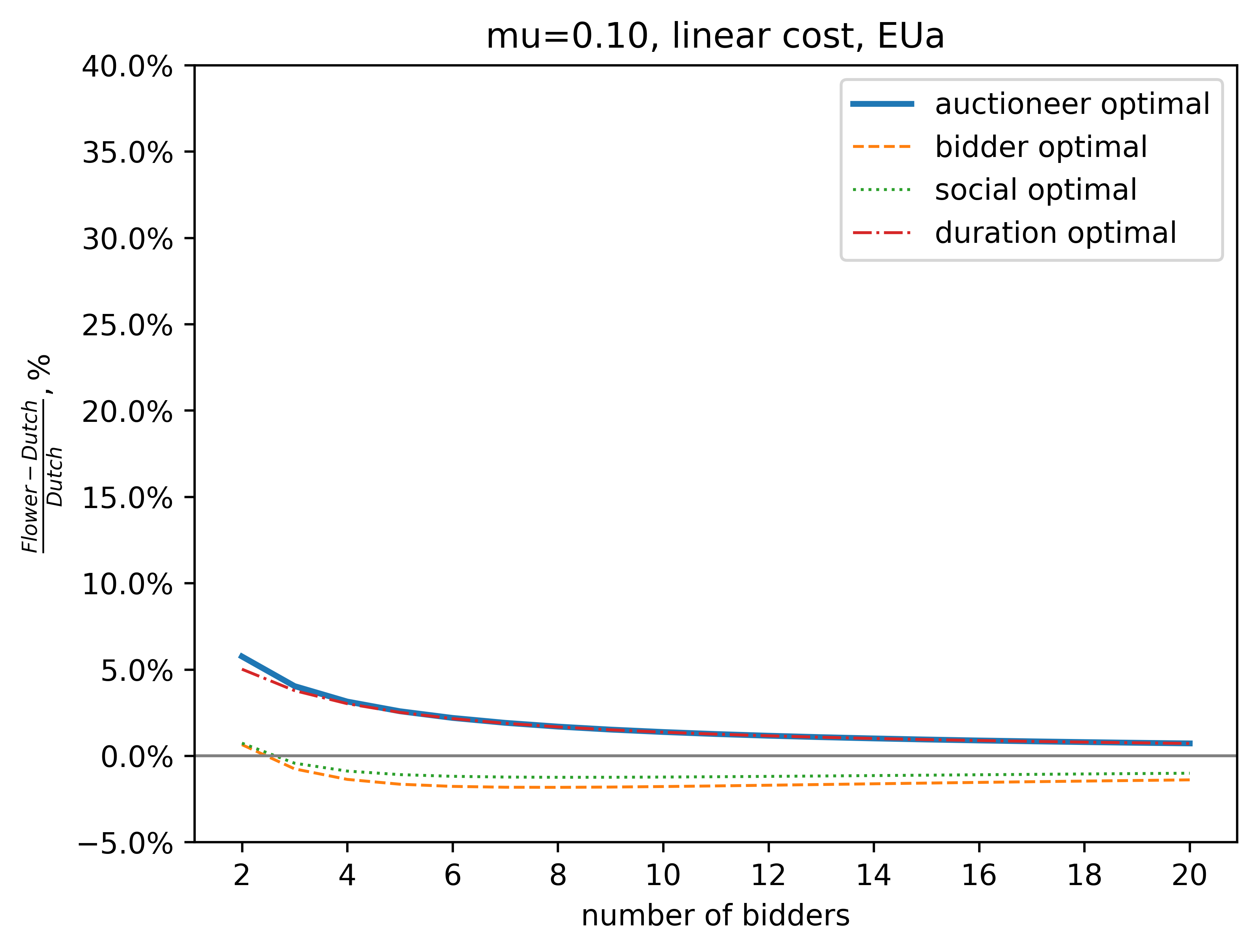}
        \caption{Low cost}
    \end{subfigure}
    \begin{subfigure}{0.48\textwidth}
        \centering
        \includegraphics[width=\linewidth]{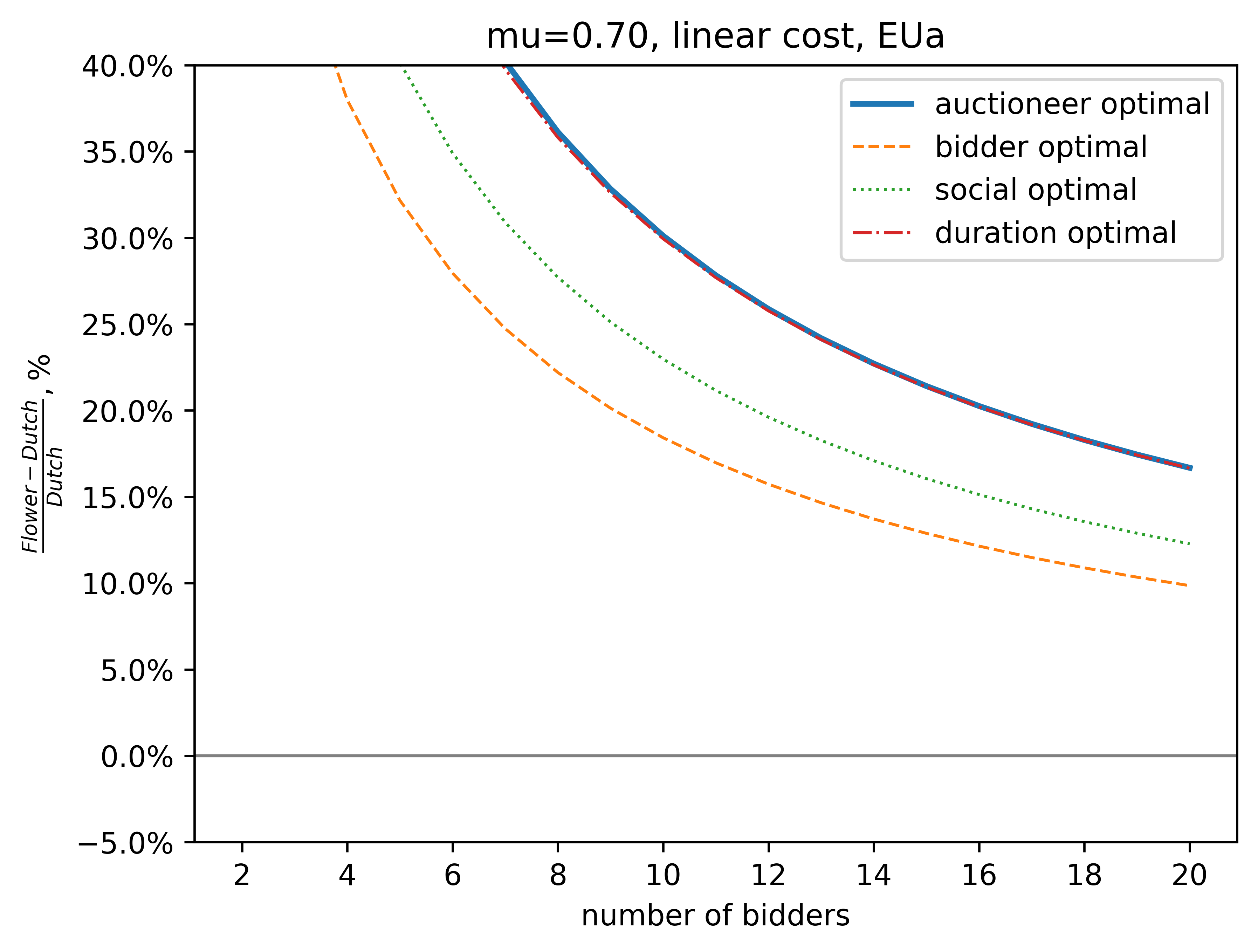}
        \caption{High cost}
    \end{subfigure}
    \centering
    \caption{Auctioneer utility under varying number of bidders}
    \label{fig:optimals_vary_n_EUa}
\end{figure}

The results of numerical simulations illustrating the advantage of the Istanbul Flower Auction relative to Dutch in terms of  auctioneer utility, calculated as $\frac{EU_{}^{F} - EU_{}^{D}}{EU_{}^{D}}$, are displayed in Figures~\ref{fig:optimals_vary_cost_EUa}
and ~\ref{fig:optimals_vary_n_EUa}. In Figure~\ref{fig:optimals_vary_cost_EUa}, we illustrate the effect of the impatience of the bidders by varying the time cost parameter from completely patient ($\mu = 0$) to highly impatient ($\mu = 0.8$) in auctions with either small number of bidders ($n = 2$) or relatively large number of bidders ($n = 10$). In Figure~\ref{fig:optimals_vary_n_EUa}, we illustrate the effect of the market competitiveness by varying the number of bidders in auctions from $2$ to a $20$ with either relatively patient bidders ($\mu = 0.1$) or highly impatient bidders ($\mu = 0.7$). 

\paragraph{Auctioneer utility at the optimal starting price}  Consider Figures~\ref{fig:optimals_vary_cost_EUa} --
 ~\ref{fig:optimals_vary_n_EUa}, and let us focus on the blue lines that represent the performance of the Istanbul Flower Auction with the auctioneer-optimal starting price. The relative advantage of the optimal Istanbul Flower Auction over the Dutch auction increases when bidders are more impatient (see Figure~\ref{fig:optimals_vary_cost_EUa}), while it decreases when the market becomes more competitive (see Figure~\ref{fig:optimals_vary_n_EUa}). When bidders are relatively patient (see panel (a) in Figure~\ref{fig:optimals_vary_n_EUa}), the optimal Istanbul Flower Auction only slightly outperforms the Dutch auction by less than $10\%$, while it still outperforms the Dutch auction by more than $15\%$ in highly competitive markets with more than 10 bidders when bidders are highly impatient (see panel (b) in Figure~\ref{fig:optimals_vary_n_EUa}).

\paragraph{Auctioneer utility at alternative starting prices} Now consider the other lines in Figure~\ref{fig:optimals_vary_cost_EUa} and Figure~\ref{fig:optimals_vary_n_EUa}, which display the auctioneer's relative benefit of the Istanbul Flower Auction over the Dutch auction when the starting price is set  to maximize the expected utility for the bidders (green lines), the expected social welfare (yellow lines) or to minimize the expected duration of the auction (red lines). We observe that the auctioneer can still significantly benefit from the Istanbul Flower Auction when bidders are very impatient. In contrast, the auctioneer can be slightly worse off (no more than $3\%$) in the Istanbul Flower Auction than in the Dutch auction when bidders are relatively patient. This suggests that the significant advantage of the Istanbul Flower Auction over the Dutch Auction is robust to alternative optimization targets. The auctioneer incurs very little loss and typically benefits a lot from the Istanbul Flower Auction compared to the Dutch auction when she prioritizes the satisfaction of bidders, the social welfare, or the speedy sale.

\paragraph{Welfare and speed performance at the auctioneer-optimal starting price} Table~\ref{table::welfare_and_duration} shows that the bidders, the social planner, and the time-saver (duration minimizer) all benefit from the auctioneer-optimal Istanbul Flower Auction compared to the Dutch auction. The relative advantage of the Istanbul Flower Auction is always greater with more impatient bidders. As the number of bidders increases, the improvements made by the Istanbul Flower Auction in terms of buyer payoff and social welfare usually become larger, except for the social welfare with highly impatient bidders.\footnote{The expected buyer payoff will be very close to zero when the number of bidders becomes large. Although the relative difference is extremely large in this case, it does not have much impact in the welfare.} 
Moreover, the speed advantage of the Istanbul Flower Auction relative to the Dutch auction becomes more pronounced in more competitive markets with highly impatient bidders, while this advantage diminishes in more competitive markets with relatively patient bidders. 

\begin{table}[!hbt] \small
    \centering
    \caption{Optimal Istanbul Flower Auction performance relative to Dutch}
    \label{table::welfare_and_duration}
    \begin{threeparttable}
    \begin{tabular} {l c c c c}
        \hline\hline
        ~ & \multicolumn{2}{c}{2 bidders} & \multicolumn{2}{c}{10 bidders} \\
        ~ & $\mu=0.1$ & $\mu=0.7$ & $\mu=0.1$ & $\mu=0.7$ \\
        \cmidrule(r){2-3}\cmidrule(r){4-5}
        Auctioneer payoff \\
        \quad Flower/Dutch, $\%$ & 105.75\% & 207.74\% & 101.38\% & 130.11\% \\
        Buyer payoff \\
        \quad Flower/Dutch, $\%$ & 107.73\% & 231.87\% & 168.19\% & 385.39\% \\
        Social welfare \\
        \quad Flower/Dutch, $\%$ & 106.73\% & 218.19\% & 108.00\% & 152.83\% \\
        Auction duration \\
        \quad Flower/Dutch, $\%$ & 18.87\% & 17.21\% & 28.85\% & 16.64\% \\
        \hline\hline
    \end{tabular}
    \end{threeparttable}
\end{table}




\section{Conclusion}\label{sec:conclusion}

Our analyses reveal  that the Istanbul Flower Auction format, with its innovative approach to combining Dutch and English auction elements, offers distinct advantages in achieving higher utilities for both auctioneers and bidders. This format proves particularly beneficial in markets characterized by high time sensitivity, such as perishable goods auctions. The flexibility to switch between different auction dynamics based on the initial bidding activity allows for more efficient price discovery and lower time costs, and  can lead to higher utilities for all parties involved. Our numerical findings illustrate that both the auctioneer and the bidders prefer the Istanbul Flower Auction over the Dutch auction, and the Istanbul Flower Auction sells goods faster than the traditional Dutch auction. This insight aligns with real-world scenarios where sellers need to sell goods swiftly. 

Our findings suggest that adopting flexible auction formats like the Istanbul Flower Auction could enhance outcomes for all the parties involved in various auction-based markets, particularly those where the speed of the auction is important due to the large number of items needing to be sold in a fixed amount of time.

\section{Declaration of competing interest}

The authors have no known competing financial interests or personal relationships that could have appeared to influence the work reported in this paper.

\section{Data availability}

No data was used for the research described in the article.

\appendix

\section*{Appendix: Proofs}\label{sec:appendix_proof}
\renewcommand{\thesubsection}{\Alph{subsection}}

\begin{proof}[Proof of Lemma~\ref{lemma::cutoff}]
    \noindent In contrast to the Dutch auction setting where the price continuously descends from $s=1$ and
    one can solve for a strictly increasing bidding function, the bidding function here may be capped by the starting price $s$ when the bidder's private value exceeds a certain cutoff $\lambda(s) \ge s$.
    Therefore, the bidding function for the subsequent Dutch auction is given by
    \begin{equation*}
        \beta(v, s) = \begin{cases}
            b(v, s) & v \le \lambda(s)          \\
            s       & \lambda(s) \le v \le p(s)
        \end{cases}
    \end{equation*}
    and the bidder's expected utility is given by
    \begin{equation*}
        EU_{B}^{FD}(v, s) = \begin{cases}
            [c(s - b(v, s))v - b(v, s)]G(v)                                                                                    & v \le \lambda(s)          \\
            \sum\limits_{k=0}^{n-1} \binom{n - 1}{k}\frac{v - s}{k + 1} F^{n - 1 - k}(\lambda(s))[F(p(s)) - F(\lambda(s))]^{k} & \lambda(s) \le v \le p(s)
        \end{cases}
    \end{equation*}
    where $b(\cdot, s) \rightarrow [0, s]$ is a strictly increasing and differentiable function with $b(0) = 0$ given the starting price $s$.

    \textbf{Step 1} We prove $\lambda(s) = p(s)$ by contradiction in two cases, depending on whether the subsequent auction could have an English phase or not.

    Let us start from the case $p(s) < 1$, where the auction could proceed with either  Dutch or English phases. Suppose $\lambda(s) < p(s)$; then in equilibrium a bidder with value $v = p(s)$ should be indifferent between bidding or not at the starting price $s$. That is,
    \begin{equation*}
        EU_{B}^{FD}(p(s), s) = EU_{B}^{FE}(p(s), s).
    \end{equation*}
    We know that
    \begin{equation*}
        EU_{B}^{FE}(p(s), s) = (p(s) - s)G(p(s))
    \end{equation*}
    and
    \begin{align*}
        EU_{B}^{FD}(p(s), s) & = \sum\limits_{k=0}^{n-1} \binom{n - 1}{k}\frac{p(s) - s}{k + 1} F^{n - 1 - k}(\lambda(s))[F(p(s)) - F(\lambda(s))]^{k} \\
                             & = \frac{p(s) - s}{n}\frac{F^{n}(p(s)) - F^{n}(\lambda(s))}{F(p(s)) - F(\lambda(s))} .
    \end{align*}
    Let us define an auxiliary function
    \begin{equation*}
        k(x) = \frac{1 - x^{n}}{n(1 - x)} = \frac{1 + x + \cdots + x^{n-1}}{n}.
    \end{equation*}
    It is obvious that $k'(x) > 0$ for $x \in (0, 1)$, and $k(1) = 1$. Note that $\frac{F(\lambda(s))}{F(p(s))} \in (F(\lambda(s)), 1)$. Then we have
    \begin{equation*}
        \frac{EU_{B}^{FD}(p(s), s)}{EU_{B}^{FE}(p(s), s)} = k\left(\frac{F(\lambda(s))}{F(p(s))}\right) < k(1) = 1.
    \end{equation*}
    which means we can never have the necessary condition for the equilibrium satisfied. Therefore, we conclude that there is no equilibrium such that $\lambda(s) < p(s)$ when $p(s) < 1$.

    Now let us look into the case $p(s) = 1$, where the Istanbul Flower Auction becomes a Dutch auction with the starting price $s$. Suppose $\lambda(s) < p(s)$. Then, any Dutch bidder whose bid is capped at $s$ will deviate to bidding at the starting price. Indeed she will win the item at the starting price for sure by turning the auction to the English phase when all other bidders wait for the Dutch phase.

    \textbf{Step 2} Given that $\lambda(s) = p(s)$ in equilibrium when $p(s) < 1$, we invoke the indifference condition at $v = p(s)$ again. We have
    \begin{align*}
        EU_{B}^{FD}(p(s), s)                        & = EU_{B}^{FE}(p(s), s) \\
        [c(s - b(p(s), s))p(s) - b(p(s), s)]G(p(s)) & = (p(s) - s)G(p(s))    \\
        c(s - b(p(s), s))p(s) - b(p(s), s)          & = p(s) - s
    \end{align*}
    By the definition of the time cost function, we have $c(0) = 1$. Thus, $b(p(s), s) = s$ is a solution, and we denote this implicitly solved cutoff by $p_0$. Since we have proved in Step 1 that the Dutch phase bidding strategy is not capped (and therefore strictly increasing from $0$ to at most $s$), for the uniqueness of the solved cutoff under a given $s$, we only need to confirm that no other $p \in [s, p_0)$ can satisfy the aforementioned indifference condition. This is equivalent to proving the uniqueness of the zero point in $[s, p_0]$ for the auxiliary function defined below:
    \begin{equation*}
        l(p) = (p - s) - [c(s - b(p, s))p - b(p, s)].
    \end{equation*}
    It is obvious that $l(p)$ is differentiable and $l(p_0) = 0$. We compute its first-order derivative and reformat it by substituting in Equation~\ref{eq::dutch_ode} evaluated at $v = p$. We have
    \begin{align*}
        l^{'}(p) & = 1 - \left[ - c'(s - b(p, s))\frac{\partial b(p, s)}{\partial p}p + c(s - b(p, s)) - \frac{\partial b(p, s)}{\partial p} \right] \\
        & = 1 - c(s - b(p, s)) + \frac{g(p)}{G(p)}[c(s - b(p, s))p - b(p, s)].
    \end{align*}
    For $p \in [s, p_0)$, by what we have proved in Step 1 for the Dutch phase bidding strategy, we have $s - b(p, s) > 0$. Then by the definition of the time cost function, we have $c(s - b(p, s)) < 1$. By individual rationality, we have $c(s - b(p, s))p - b(p, s) \ge 0$. Therefore, we have $l^{'}(p) > 0$. Now we know that $l(p)$ is strictly increasing in $(s, p_0)$ with $l(p_0) = 0$, which means it has a unique zero point $p_0$ in $[s, p_0]$.

    When $p(s) = 1$, we do not necessarily have $b(p, s) = s$ since the starting price $s$ may be too high to put any restriction on any bidder's Dutch phase decision. However,
    it must be true that $b(p, s) \leq s.$
\end{proof}

\begin{proof}[Proof of Lemma~\ref{lemma::partial_b_s_upperbound}]
    Taking the partial derivative with respect to $s$ on both sides of equation~\ref{eq::dutch_ode}, we have
    \begin{equation*}
        \frac{\partial^2 b(v, s)}{\partial s \partial v} = \frac{g(v)}{G(v)}\frac{\begin{matrix}
                \left[c'(s - b(v, s))\left(1 - \frac{\partial b(v, s)}{\partial s}\right)v - \frac{\partial b(v, s)}{\partial s}\right][1 + c'(s - b(v, s))v] & \\
                - [c(s - b(v, s))v - b(v, s)]c''(s - b(v, s))\left(1 - \frac{\partial b(v, s)}{\partial s}\right)v
            \end{matrix}}{[1 + c'(s - b(v, s))v]^2}.
    \end{equation*}
    The above equation can be viewed as a differential equation of the function $\frac{\partial b(v, s)}{\partial v}$ 
    and its derivative with respect to $s$. Since we focus on the well-behaved increasing bidding function $b(v, s)$ that allows us to interchange the order of partial derivatives such that $\frac{\partial^2 b(v, s)}{\partial s \partial v} = \frac{\partial^2 b(v, s)}{\partial v \partial s}$. Note that We also have the initial condition $b(0)\equiv 0$, which gives the initial condition $\frac{\partial b(v, s)}{\partial s}|_{v=0} = 0$ for the above differential equation. Following the path of $\frac{\partial b(v, s)}{\partial s}$, when the value of the function $\frac{\partial b(v, s)}{\partial s} \rightarrow 1$ at certain $v$, its derivative $\frac{\partial}{\partial v}\frac{\partial b(v, s)}{\partial s} < 0$ at that point. Therefore, starting from the value of $\frac{\partial b(0, s)}{\partial s}\equiv 0$, the function $\frac{\partial b(v, s)}{\partial s}$ is always below the upper bound that $\frac{\partial b(v, s)}{\partial s} <1$.
\end{proof}

\begin{proof}[Proof of Lemma~\ref{lemma::partial_m_s_bound}]
    By definition, we have
    \begin{equation*}
        c(m(v, s) - s)v - m(v, s) = 0.
    \end{equation*}
    When there is no time cost, $c(t) = 1$, we have the classical result $m(v, s) = v$.

    Otherwise, we must have $c^{'}(t) < 0$. Taking the derivative with respect to $s$ on both sides, we have
    \begin{equation*}
        c'(m(v, s) - s)\left( \frac{\partial m(v, s)}{\partial s} - 1 \right) v - \frac{\partial m(v, s)}{\partial s} = 0
    \end{equation*}
    which gives
    \begin{equation*}
        \frac{\partial m(v, s)}{\partial s} = - \frac{c'(m(v, s) - s)v}{1 - c'(m(v, s) - s)v} \in (0, 1)
    \end{equation*}
\end{proof}

\begin{proof}[Proof of Lemma~\ref{lemma::increasing_cutoff}]
    The starting price $s$ puts a cap on the Dutch phase bids. By Lemma~\ref{lemma::partial_b_s_upperbound}, we know that the Dutch phase bidding function will never move upward faster than the cap $s$. It is also evident that $b(1, s) < 1$. Therefore, when  $s$ increases from $0$ to $1$, it departs upwards from the Dutch phase bidding function and finally becomes non-restrictive to the Dutch phase bidding strategy after reaching some interim $\tilde{s}$. 
    By Lemma~\ref{lemma::cutoff}, we can solve for $\tilde{s}$ from $b(1, \tilde{s}) = \tilde{s}$, and use the monotonicity of the $l(\cdot)$ function defined in the proof of that lemma to prove that $p(s) \equiv 1$ for all $s \in [\tilde{s}, 1]$.

    Now, let us consider $s \in [0, \tilde{s})$ where the Dutch phase bids of high-value bidders are capped. By Lemma~\ref{lemma::cutoff} we know that
    \begin{equation*}
        b(p(s), s) = s.
    \end{equation*}
    Taking the derivative with respect to $s$ on both sides of the above equation, we have
    \begin{equation*}
        \left. \frac{\partial b(v, s)}{\partial v}  \right|_{v=p(s)} \frac{dp(s)}{ds} + \left. \frac{\partial b(v, s)}{\partial s} \right|_{v=p(s)} = 1.
    \end{equation*}
    We are considering strictly increasing Dutch bidding strategies, which means $\frac{\partial b(v, s)}{\partial v} > 0$. By Lemma~\ref{lemma::partial_b_s_upperbound}, we have $\frac{\partial b(v, s)}{\partial s} < 1$. Therefore, we conclude that
    \begin{equation*}
        \frac{dp(s)}{ds} > 0.
    \end{equation*}
\end{proof}

\begin{proof}[Proof of Lemma~\ref{lemma::dbds}]
    From the initial condition of the Dutch phase differential equation~\ref{eq::dutch_ode}, $b(0, s) = 0$ as the bidder with zero value always bids zero, we obtain that $\frac{\partial b(0, s)}{\partial s} = 0$. 

    Suppose that there exists $s_1 < s_2, v_0 \in (0, 1]$ such that $b(v_0, s_1) = b(v_0, s_2) = b_0$, from the Dutch phase differential equation~\ref{eq::dutch_ode} we have
    \begin{equation*}
        \frac{\partial b(v_0, s_1)}{\partial v}\frac{1 + v_0c'(s_1 - b_0)}{v_0c(s_1 - b_0) - b_0} = \frac{\partial b(v_0, s_2)}{\partial v}\frac{1 + v_0c'(s_2 - b_0)}{v_0c(s_2 - b_0) - b_0}.
    \end{equation*}
    Under the assumption, we know that $c'(s_1 - b_0) \le c'(s_2 - b_0) < 0$, while the equal sign only holds for the linear cost. We also have, by definition of the cost function, $c(s_1 - b_0) > c(s_2 - b_0) > 0$. Then we must have
    \begin{equation*}
        \frac{\partial b(v_0, s_1)}{\partial v} > \frac{\partial b(v_0, s_2)}{\partial v}.
    \end{equation*}
    For $h \rightarrow 0$, applying Euler's method to the differential equation with initial condition $b(0, s) = 0$ to get
    \begin{equation*}
        b(h, s) \approx b(0, s) + h\frac{\partial b(0, s)}{\partial v} = 0,
    \end{equation*}
    and substitute this into the Dutch differential equation 
    to get
    \begin{equation*}
        \frac{\partial b(h, s)}{\partial v} \approx \frac{g(h)}{G(h)}\frac{hc(s)}{1 + hc'(s)}.
    \end{equation*}
    Again, given that $s_1 < s_2$, we must have $c(s_1) > c(s_2) > 0$ by definition of the cost function, and under the assumption we have $c'(s_1) \le c'(s_2) < 0$. Then we have
    \begin{equation*}
        \frac{\partial b(0, s_1)}{\partial v} > \frac{\partial b(0, s_2)}{\partial v}.
    \end{equation*}
    Now we know that, for any starting price $s_1 < s_2$, the two Dutch phase bidding functions $b(v, s_1)$ and $b(v, s_2)$ have the following properties: (i) the two functions share the same starting point $b(0, s) = 0$; (ii) $b(v, s_1)$ is steeper than $b(v, s_2)$ at $v = 0$; (iii) whenever $b(v, s_1)$ and $b(v, s_2)$ have a non-zero intersection point, $b(v, s_1)$ is steeper than $b(v, s_2)$ at that point. Suppose such non-zero intersection points exist, we must be able to find the smallest among them, denoted by $v_m > 0$. By the aforementioned properties, there exists $\varepsilon_1, \varepsilon_2 > 0$ such that $b(v_m - \varepsilon_1, s_2) > b(v_m - \varepsilon_1, s_1), b(\varepsilon_2, s_2) < b(\varepsilon_2, s_1)$. Then, according to the intermediate value theorem, there must be another intersection point between $0$ and $v_m$, which gives a contradiction. Therefore, we conclude that
    \begin{equation*}
        \frac{\partial b(v, s)}{\partial s} < 0, \forall v \in (0, 1].
    \end{equation*}

\end{proof}

\begin{proof}[Proof of Proposition~\ref{prop::flower_better}]
    By Lemma~\ref{lemma::cutoff} and Lemma~\ref{lemma::increasing_cutoff}, we have either
    \begin{equation*}
        s \in [0, \tilde{s}), p(s) < 1, \frac{dp(s)}{ds} > 0, b(p(s), s) = s
    \end{equation*}
    or  
    \begin{equation*}
        s \in (\tilde{s}, 1], p(s) = 1, \frac{dp(s)}{ds} = 0, b(p(s), s) < s.
    \end{equation*}
    Specifically, $p(s)$ is not differentiable at $\tilde{s}$: the left-sided limit belongs to the first case, and the right-sided limit belongs to the second case. However, plugging these into equation~\ref{eq::dEUa} indicates that $EU_{A}^{F}(s)$ is differentiable at that point, and the first-order derivative of the auctioneer utility for $s \in [\tilde{s}, 1)$ is given by
                \begin{equation*}
                    \frac{dEU_{A}^{F}(s)}{ds} = \int_{0}^{1} \frac{\partial b(v, s)}{\partial s} \, dF^{n}(v) < 0
                \end{equation*}
                because by Lemma~\ref{lemma::dbds} we have $\frac{\partial b(v, s)}{\partial s} < 0$ for all $v \in (0, 1]$. Therefore, we must have $EU_{A}^{F}(s^*) > EU_{A}^{F}(s) \ge EU_{A}^{F}(1) = EU_{A}^{D}$ for any $s \in [\tilde{s}, 1]$. It is obvious that $s^* < \tilde{s}$.

    Using equation~\ref{eq::dEUa} again, we can compute that
    \begin{equation*}
        \frac{dEU_{A}^{F}(0)}{ds} = - \int_{0}^{1} \int_{0}^{v} \frac{\partial m(v, s)}{\partial s}h(v, x) \, dx \, dv < 0
    \end{equation*}
    because by Lemma~\ref{lemma::partial_m_s_bound} we have $\frac{\partial m(v, s)}{\partial s} \in (0, 1)$. Therefore, we must have $EU_{A}^{F}(s^*) > EU_{A}^{F}(0) = EU_{A}^{D}$ for any $s \in [\tilde{s}, 1]$, which also implies $s^* > 0$.

    Finally, we can conclude that $s^* \in (0, \tilde{s})$,  
    and therefore the optimal auction is non-trivial.
\end{proof}

\begin{proof}[Proof of Proposition~\ref{prop::flower_bidder_better}]
    Below we show that the first-order derivative of the expected utility for bidders is (1) increasing at $s = 0$ so there exists a better Istanbul Flower Auction than the standard English auction, and (2) decreasing in $(\tilde{s}, 1)$ so there exists a better Istanbul Flower Auction than the standard Dutch auction. Given the obvious continuity of the utility function, the proposition then follows naturally.

    \textbf{(1)} The first-order derivative of the bidder utility at $s = 0$ is given by
    \begin{equation*}
        \frac{dEU_{B}^{F}(0)}{ds} = \int_{0}^{1} \int_{0}^{v} \left[
            c^{'}(m(x, 0))\left( \frac{dm(x, 0)}{ds} - 1 \right) v - \frac{dm(x, 0)}{ds}
        \right] \, dG(x) \, dF(v).
    \end{equation*}
    Taking the derivative with respect to $s$ on both sides of Equation~\ref{eq::english_strategy} we have
    \begin{equation*}
        c^{'}(m(v, s))\left( \frac{dm(v, s)}{ds} - 1 \right) v - \frac{dm(v, s)}{ds} = 0
    \end{equation*}
    which implies
    \begin{equation*}
        c^{'}(m(x, s))\left( \frac{dm(x, s)}{ds} - 1 \right) v - \frac{dm(x, s)}{ds} \ge 0
    \end{equation*}
    where the equality holds only for $x = v$, because by Assumption~\ref{assumption::cost} we have $c^{'}(t) < 0$, by Lemma~\ref{lemma::partial_m_s_bound} we have $\frac{dm(x, s)}{ds} < 1$, and by our model setting $x$ is the second highest value which must be less than or equal to $v$. Therefore, considering the integrand of the derivative, we have that the strict inequality holds, that is $\frac{dEU_{B}^{F}(0)}{ds} > 0$.

    \textbf{(2)} The first-order derivative of the bidder utility for $s \in [\tilde{s}, 1)$ is given by
    \begin{eqnarray*}
        \frac{dEU_{B}^{F}(s)}{ds} &=& \int_{0}^{1} \left( c'(s-b(v, s))v-[1+c'(s-b(v, s))v] \frac{\partial b(v, s)}{\partial s} \right) G(v) \, dF(v)
    \end{eqnarray*}
    Let us denote the integrand by
    \begin{equation*}
        w(v, s) = G(v)\left(c'(s-b(v, s))v-[1+c'(s-b(v, s))v] \frac{\partial b(v, s)}{\partial s}\right)
    \end{equation*}
    By the initial condition $b(0) = 0$ for the Dutch phase differential equation, we know that $\frac{\partial b(0, s)}{\partial s} = 0$ and then we have $w(0, s) = 0$. Taking the derivative with respect to $v$, we have
    \begin{eqnarray*}
        \frac{\partial  w(v, s)}{\partial v} &=& G(v)\left(\left[-c''(s-b(v,s))\frac{\partial b(v, s)}{\partial v}v + c'(s-b(v,s))\right] \left( 1 - \frac{\partial b(v, s)}{\partial s} \right) \right. \\
        & & \left. - [1+c'(s-b(v,s)) v]  \frac{\partial ^2 b(v,s)}{\partial s \partial v}\right)
        \\
        & &+ g(v)\left( c'(s-b(v, s))v-[1+c'(s-b(v, s))v] \frac{\partial b(v, s)}{\partial s} \right) \\
    \end{eqnarray*}
    Substitute in $\frac{\partial^2 b(v, s)}{\partial v \partial s}$,
    \begin{eqnarray*}
        \frac{\partial^2 b(v, s)}{\partial v \partial s} &=& \frac{g(v)}{G(v)}\frac{
            \left[c'(s - b(v, s))\left(1 - \frac{\partial b(v, s)}{\partial s}\right)v - \frac{\partial b(v, s)}{\partial s}\right][1 + c'(s - b(v, s))v]
        }{[1 + c'(s - b(v, s))v]^2} \\
        && - \frac{g(v)}{G(v)}\frac{[c(s - b(v, s))v - b(v, s)]c''(s - b(v, s))\left(1 - \frac{\partial b(v, s)}{\partial s}\right)v}{[1 + c'(s - b(v, s))v]^2}
    \end{eqnarray*}
    we have $\frac{\partial  w(v, s)}{\partial v}$ simplified as:
    \begin{eqnarray*}
        \frac{\partial  w(v, s)}{\partial v} &=& G(v)\left[-c''(s-b(v,s))\frac{\partial b(v, s)}{\partial v}v + c'(s-b(v,s))\right] \left( 1 - \frac{\partial b(v, s)}{\partial s} \right) \\
        && + g(v)\frac{(c(s-b(v))v-b)c''(s-b(v,s))(1- \frac{\partial b(v, s)}{\partial s})v}{1 + c'(s - b(v, s))v}
    \end{eqnarray*}
    Substitute in $\frac{\partial b(v, s)}{\partial v}$ in (\ref{eq::dutch_ode}):
    \begin{equation*}
        \frac{\partial  w(v, s)}{\partial v}  =G(v)c'(s-b(v,s))\left( 1- \frac{\partial b(v, s)}{\partial s} \right)
    \end{equation*}
    By Assumption~\ref{assumption::cost} and Lemma \ref{lemma::partial_b_s_upperbound}, we know that $c'(t)<0$ and $1- \frac{\partial b(v, s)}{\partial s}>0$. Therefore, $\forall v\in(0,1),\;w(v,s)$ is decreasing in $v$ with $s\in (\tilde{s}, 1)$.
    Combining with $w(0,s)\equiv 0$ as the bidder with zero value always gets zero utility, we have $\frac{dEU_{B}^{F}(s)}{ds} < 0$ for $s \in (\tilde{s}, 1)$.
\end{proof}

\begin{proof}[Proof of Proposition~\ref{prop::social_welfare}]
    Differentiating Equation~\ref{eq::EUab} with respect to the starting price $s$, for $s \in (\tilde{s}, 1)$, we have
    \begin{eqnarray*}
        \frac{\partial EU_{S}^{F}(s)}{\partial s} &=& \int _{0}^{1} c'(s-b(v, s)) \left( 1-\frac{\partial b(v, s)}{\partial s} \right) v \, dF^n(v) < 0
    \end{eqnarray*}
    and for $s = 0$, we have
    \begin{eqnarray*}
        \frac{\partial EU_{S}^{F}(s)}{\partial s} &=& \int _{0}^{1} \int _{0}^{v}  c'(m(v, s)-s) \left( \frac{\partial m(v, s)}{\partial s} -1 \right) vh(v, x) \, dx \, dv > 0
    \end{eqnarray*}
    since both $\frac{\partial b(v, s)}{\partial s} < 1$ and $\frac{\partial m(v, s)}{\partial s} < 1$. The proposition directly follows.
\end{proof}

\begin{proof}[Proof of Proposition~\ref{prop::time-saving}]
    Differentiating Equation~\ref{eq::ED} with respect to the starting price $s$, for $s \in (\tilde{s}, 1)$, we have
    \begin{equation*} \frac{\partial ED^{F}(s)}{\partial s} = \int _{0}^{1} \left( 1-\frac{\partial b(v, s)}{\partial s} \right)\, dF^n(v) > 0
    \end{equation*}
    and for $s=0$, we have
    \begin{equation*}
        \frac{\partial ED^{F}(s)}{\partial s} = \int _{0}^{1}\int_{0} ^{v} \left( \frac{\partial m(v, s)}{\partial s} - 1 \right) h(v,x)\, dx  \, dv <0
    \end{equation*}
   since  both $\frac{\partial b(v, s)}{\partial s} < 1$ and $\frac{\partial m(v, s)}{\partial s} < 1$. The proposition directly follows.
\end{proof}

\bibliographystyle{elsarticle-harv}

\end{document}